\documentclass[mathleft
]{an}
\usepackage{graphicx}
\usepackage{times}
\begin{document}

\Pagespan{001}{}
\Yearpublication{????}%
\Yearsubmission{2012}%
\Month{12}%
\Volume{???}%
\Issue{??}%

\title{Discussing the physical meaning of the absorption
        feature at 2.1 keV in 4U 1538$-$52\,\thanks{Data
from XMM-Newton}}

\author{J.J. Rodes-Roca\inst{1,2}\fnmsep\thanks{Corresponding author:
  \email{rodes@dfists.ua.es}\newline}
\and  J.M. Torrej\'on\inst{1,2} \and S. Mart\'{\i}nez-Nu\~nez\inst{2}
\and A. Gim\'enez-Garc\'{\i}a\inst{1,2} \and G. Bernab\'eu\inst{1,2}
}
\titlerunning{Discussing the physical meaning of the 2.1 keV feature}
\authorrunning{J.J. Rodes-Roca et al.}
\institute{
Department of Physics, Systems Engineering and Signal Theory
University of Alicante, 03080 Alicante, Spain
\and 
University Institute of Physics Applied to Sciences and Technologies,
University of Alicante, 03080 Alicante, Spain}

\received{30 Dec 2012}
\accepted{?? ??? 20??}
\publonline{later}


\abstract{%
High resolution X-ray spectroscopy is a powerful tool for
studying the nature of the matter surrounding the neutron star in
X-ray binaries and its interaction between the stellar wind and the
compact object. In particular, absorption features in their
spectra could reveal the presence of atmospheres of the
neutron star or their magnetic field strength.
Here we present an investigation of the absorption feature at
2.1 keV in the X-ray spectrum of the high mass X-ray
binary 4U 1538$-$52 based on our previous analysis of the
XMM-Newton data. We study various possible origins
and discuss the different physical scenarios
in order to explain this feature. A likely interpretation is
that the feature is associated with atomic transitions in an
O/Ne neutron star atmosphere or of hydrogen and helium like
Fe or Si ions formed in the stellar wind of the donor.}

\maketitle

\section{Introduction}
\label{intro}

The improvement of the capabilities of modern X-ray observatories,
like {\it Chandra} or {\it XMM-Newton}, offers the possibility
to detect and analyse both absorption and emission lines and to study the
nature of the matter surrounding the compact object in many X-ray sources.
The presence of emission lines in high mass X-ray binary systems (HMXBs) has been reported since the beginning of X-ray astronomy
with different X-ray observatories, e. g., Centaurus X-3 with \emph{ASCA} and
\emph{Chandra} (\cite{ebisawa96}; \cite{iaria05}), Vela X-1 with \emph{ASCA} (\cite{sako99}),
4U 1700$-$37 with \emph{XMM-Newton} (\cite{vdMeer05}),
Cygnus X-3 with \emph{Chandra} (\cite{paerels2000}), and 4U 1538$-$52 with \emph{XMM-Newton}
(\cite{RR11}, hereafter referred to as Paper I).

The presence of absorption lines in HMXBs is scarcer and has been associated
with cyclotron resonant scattering features (CRSFs) at energies greater than 10 keV, e.g.,
4U 1907$+$09 with \emph{BeppoSAX} (\cite{cusumano98}), 4U 1538$-$52 with \emph{Ginga},
\emph{RXTE} and \emph{INTEGRAL} (\cite{clark90}; \cite{rodesPhD}; \cite{jjrr09}), Vela X-1
with \emph{Mir-HEXE}
and \emph{RXTE} (\cite{kendziorra92}; \cite{ingo2002}; \cite{ingophd}), Centaurus X-3 with
\emph{BeppoSAX} (\cite{santangelo98}; \cite{HCh99}), OAO 1657$-$415 with \emph{BeppoSAX}
(\cite{orlandini99}), GX 301$-$2 with \emph{Ginga} (\cite{mihara95}), and LMC X-4 with
\emph{BeppoSAX} (\cite{labarbera2001}). Vela X-1 is the only HMXB
where absorption K-edges, such as O, Si, S, and K- and L-edges of Fe have been reported at energies below 10 keV (\cite{goldstein04}).

In low mass X-ray binary systems (LMXBs) absorption lines have been detected
with both \emph{Chandra} and \emph{XMM-Newton} below 10 keV, e.g., Circinus X-1
(\cite{BS2000}; \cite{dai2007}), MXB 1659$-$298 (\cite{sidoli01}),
4U 1820$-$30 (\cite{boirin04}), and XB 1916$-$053 (\cite{cackett08}).
Other astrophysical X-ray sources, like active galactic nuclei (AGNs),
have also shown X-ray absorption lines due to both transition H-like and He-like
ions and inner-shell transitions in lower ionization species (\cite{hullac}).

It is relevant for this work to mention
the isolated neutron star (INS) 1E1207.4$-$5209, which is a unique system
among the INSs because it shows more than one absorption spectral feature in
the 0.5--3.0 keV energy range (\cite{bignami} 2003). In fact, one,
two or three of the absorption features appear above 1 keV, while
the other INSs have absorption features at $E=0.2-0.7$ keV
(\cite{hoetal07}).
There are several studies suggesting that the absorption feature at 2.1 keV in this source
corresponds to a cyclotron absorption line (the second harmonic). Discarding
the instrumental origin, \cite{bignami} (2003) explained
the 2.1 keV feature as a cyclotron resonant absorption line
because both the equivalent width and deviations from the
continuum model were much stronger that instrumental or calibration effects.
Later, \cite{mori05} concluded that
the residuals around 2.1 keV were consistent in strength and position with the
instrumental Au edge, doing a detailed analysis of the \emph{XMM-Newton} data.
However, \cite{liu06}
also conclude the feature at 2.1 keV is more likely a cyclotron absorption line.
The \emph{Suzaku} observation of this source are likely in favour of the electronic
cyclotron line (\cite{takahashi}). In conclusion, although
the interpretation of the absorption feature at 2.1 keV in this isolated neutron
star has been controversial, the cyclotron resonant scattering 
feature explanation is preferred.

The continuum models used in Paper I fit the spectrum of
4U 1538$-$52 satisfactorily. However, they failed to fit correctly the region
around 2.1 keV which shows clearly an absorption feature near 2.1 keV
(see Figure~\ref{paperI}).
In our analysis we discarded a gain instrumental effect.
It is well known the existence of edges of instrumental
origin due to the Au M edge near the energy of this feature.
Therefore we have to analyse whether the residuals are
consistent in strength and position with the Au M residuals
observed in cross-calibration sources.

In this work, we discuss
the different physical scenarios in which the detected absorption
feature could be formed. In order
to ensure that the absorption feature is not due to
an instrumental effect, we have carried
out a systematic study to rule this possibility out.
First, we have discarded completely the instrumental origin
in Section~\ref{instrumental}. Then,
we have looked
for a possible astrophysical origin (background, dust scattered halo and
source) in Section~\ref{astrorigin}. Finally, we have discussed where it could be formed either in the atmosphere of the neutron star (Section~\ref{ns}) or
in the stellar wind (Section~\ref{wind}).

\section{Observation and data analysis}

The observation of 4U 1538$-$52 was carried out using
the European Photon Imaging Camera (\emph{EPIC}) aboard the \emph{XMM-Newton} satellite.
This source was observed for $\sim$55 ks on
2003 August 14--15 (Obs. ID 0152780201). Both the \emph{EPIC/metal-oxide semiconductor (MOS)}
and \emph{EPIC/PN} instruments (\cite{struder}; \cite{turner}) were operated
in \emph{Full frame} mode, and the thin filter $1$
(\emph{MOS-1} and \emph{PN}) and medium filter (\emph{MOS-2}) were used.
The observations details were summarised in Paper I.

The \emph{EPIC/PN} observation data files (ODF) were processed using Science Analysis System (SAS) version 12.0.1 together with the
latest calibration files, CCFs as of
2012 June starting from the ODF level running \emph{epproc} and
following the standard procedure for \emph{XMM-Newton} spectra.
Particular care was done to the selection of the background,
halo, and source regions. In the \emph{PN} field of view we
selected the source and the background regions from the same chip
with a circle of 35$^{\prime\prime}$ and 60$^{\prime\prime}$ radius, respectively,
and rejecting the area possibly contaminated by out-of-time events
or too near to the CCD edges. We defined the halo with
an annulus region between 95$^{\prime\prime}$ and
137.5$^{\prime\prime}$ radius, excluding both all the point sources
and the CCD edges. The spectra were rebinned in order to have
at least 20 counts per channel.

\begin{figure}[h!t]
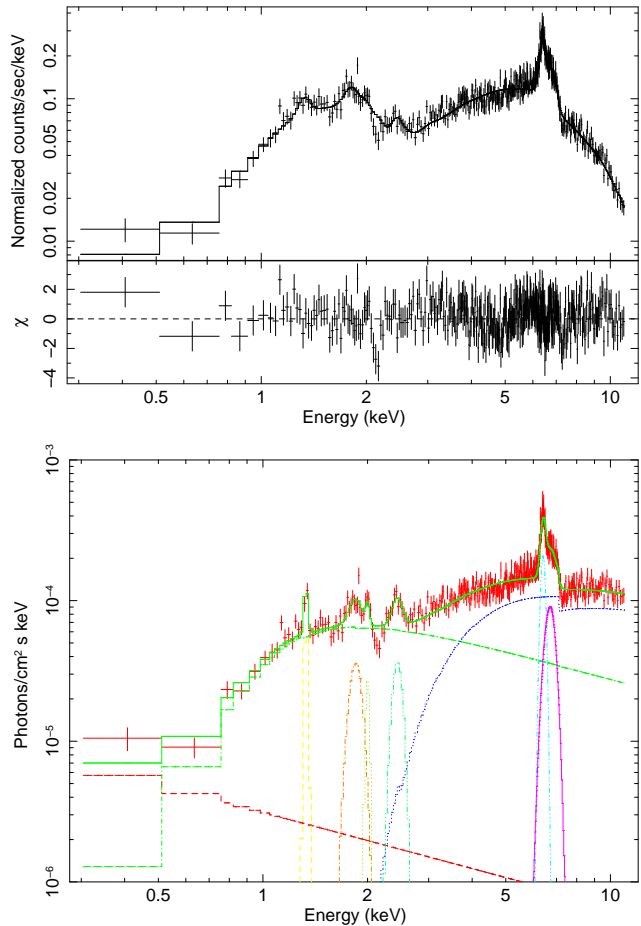

  \centering
  \includegraphics[angle=-90,width=\columnwidth]{pn_spectrum.ps}
  \includegraphics[angle=-90,width=\columnwidth]{pn_unfolded.ps}
  \caption{Iron emission lines and recombination emission
    lines of He- and H-like species
    in the eclipse spectrum of 4U 1538$-$52.
    Spectrum and best fit model (two absorbed power laws modified
    by six Gaussian emission lines) in the 0.3--11.5 keV energy range
    obtained with \emph{PN} camera and the
    residuals between the spectrum and the model (\emph{top panel}).
    Unfolded spectrum with the individual model components (\emph{bottom panel}).
    The absorption line wich is not included in the model is clearly seen.
}
  \label{paperI}
\end{figure}

The 80 ks long \emph{XMM-Newton} observation has been
divided into three time intervals. We have called the first $\approx$10 ks
as out-of-eclipse observation, from  $\approx$10 ks to $\approx$20 ks
as eclipse ingress observation, and the last $\approx$60 ks as
eclipse observation. Following our spectral analysis
in Paper I, we have modelled the X-ray continuum of the
different time intervals of the observation by using either three absorbed
power laws (out-of-eclipse and eclipse ingress observation) or two
absorbed power laws (eclipse observation). Furthermore,
the spectra of the source shows the presence of six emission
lines corresponding to iron emission lines and He and H
recombination lines (see Figure~\ref{paperI}). These lines
have been modelled by Gaussian functions. The
absorption feature at 2.1 keV presents in our
residuals has been modelled by an absorption
Gaussian function.

The spectral analysis was performed using \textsc{XSPEC}
v12.8.0 in the energy range 0.3--10.0 keV. As a first step,
as a result of the improvement of the instrumental response,
we addressed the pulse-phase averaged spectrum with respect
to our previous analysis (Paper I)
and confirmed that the spectral parameters were consistent taking
their associated errors into account. Then we have changed
this model slightly.
The Tuebingen-Boulder ISM absorption model calculates the
cross section for X-ray absorption by the ISM using
the most up-to-dates ISM abundances (\emph{tbabs} model in
\emph{XSPEC}). Therefore, we have used it instead of the
\emph{phabs} model in
\emph{XSPEC},
the absorption cross sections are adapted from \cite{verner96}
(instead of \cite{bcmc}),
and the abundances are set to those of \cite{wilms00}
(instead of \cite{angr}). We also notice that the
out-of-eclipse spectrum in Paper I has been divided into
two spectra in this work (out-of-eclipse and eclipse ingress).

\subsection{Discarding the instrumental effect}
\label{instrumental}

First of all, in order to rule out the possibility of a calibration
issue, a careful study of
this observation as well as a sample of observations used by the \emph{XMM-Newton}
cross calibration (\emph{XCAL}) archive was performed by the
\emph{XMM-Newton} calibration team.
The cross-calibration \emph{XMM-Newton} database consists of
$\sim$150 observations of different sources, optimally reduced, fitted with
spectral models defined on a source-by-source
basis\footnote{http://www.iachec.org/meetings/2009/Guainazzi\_2.pdf}.
Table~\ref{xcal} shows the EW of an unresolved
Gaussian absorption feature with centroid energy fixed at 2.1 keV in a
sample of \emph{XCAL} \emph{EPIC/PN} spectra of the sources listed below.
The \emph{XCAL} observations of
3C111, 1H1219+301, H1426+428 and PKS0548$-$322 showed no feature at 2.1 keV
(equivalent width less than 5.3 eV), while in our
observation there was clearly a feature with an EW = $-$39$^{+11}_{-21}$ eV, using our
best-fit model.
We also notice that
for on axis sources systematic calibration uncertainties
are better than 5\% in the determination of the total
effective area over the spectral range from 0.4--12.0 keV
for each \emph{EPIC} instrument separately\footnote{\emph{EPIC}
status calibration and data analysis, document
XMM-SOCCAL-TN-0018 edited by M. Guainazzi, on behalf of
the \emph{EPIC} consortium}.
The conclusion was that this absorption feature is larger than
the typical systematic uncertainties in this energy range and, therefore, it is
intrinsic to the system.
Moreover, as the equivalent width of the 2.1 keV feature
is $\sim$8 times higher than the calibration uncertainties
and the deviations from the continuum model are $\sim$4 times higher
than the calibration accuracy we can conclude that the line is resolved.

\begin{table}[htb]
  \caption{EWs in a sample of \emph{XCAL EPIC/PN} spectra.
  }
\label{xcal}
\centering                          
\begin{tabular}{l c}        
\hline                 
ObsID & EW (eV) \\    
\hline                        
0065940101 & $-5^{+3}_{-4}$ \\      
0552180101 & $-1.0^{+0.0}_{-2.1}$ \\
0111840101 & $-1^+{0}_{-7}$ \\      
0111850201 & $-0.0^{+0.0}_{-2.4}$ \\
0165770101 & $-0.0^{-0.0}_{-1.4}$ \\
0165770201 & $-1^{+0}_{-3}$ \\
0212090201 & $-0.0^{+0.0}_{-1.5}$ \\
0310190101 & $-0.0^{+0.0}_{-1.1}$ \\
0310190201 & $-2^{+0}_{-3}$ \\
0310190501 & $-0^{+0}_{-3}$ \\
0142270101 & $-0^{+0}_{-3}$ \\
\hline                                   
\end{tabular}
\end{table}

Finally, we also checked that no spectral feature at this energy is
seen in another object presented in the
same observation. We extracted the spectrum of the source
2XMM J154305.5$-$522709 (\cite{watson})
in the 0.3--10.0 keV band using the method described in Paper I.
Then we fitted the spectrum with an absorbed power-law model adding
two Gaussians for emission iron lines. The residuals between
the data and the model showed no evidence of an absorption feature
at 2.1 keV.
These results give us confidence about the
non-instrumental nature of the absorption feature.

\begin{figure}[h!t]
  \centering
  \includegraphics[angle=-90,width=\columnwidth]{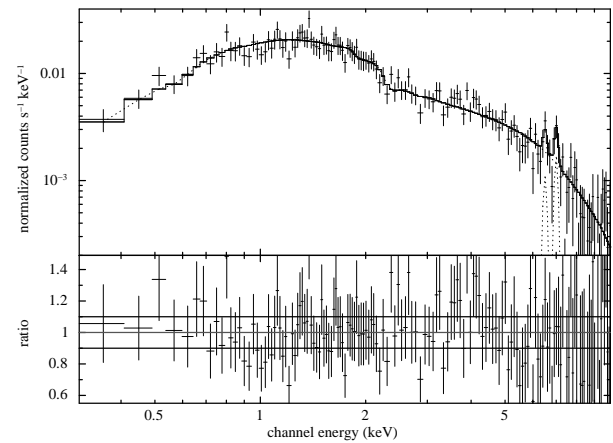}
\vspace{0.1cm}
  \caption{2XMM J154305.5$-$522709 spectrum. The lower panel
  shows the residuals between the spectrum and the best fit
  model. Horizontal lines mark the 10\% ratio differences.
    }
  \label{atmosphere}
\end{figure}
%

\subsection{Looking for the astrophysical origin}
\label{astrorigin}

\begin{figure}[h!t]
  \centering
  \includegraphics[angle=0,width=\columnwidth]{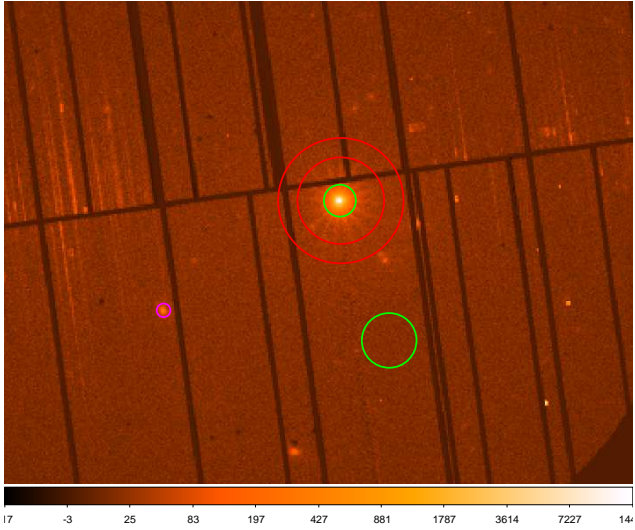}
\vspace{0.1cm}
  \caption{\emph{PN} image around 4U 1538$-$52. The open circles
  identify regions for source (middle green circle), background
  (bottom green circle), dust scattered
  halo (red annulus around central source), and 2XMM J154305.5$-$522709
  (left magenta circle).
    }
  \label{PNim}
\end{figure}

\begin{figure*}[h!t]
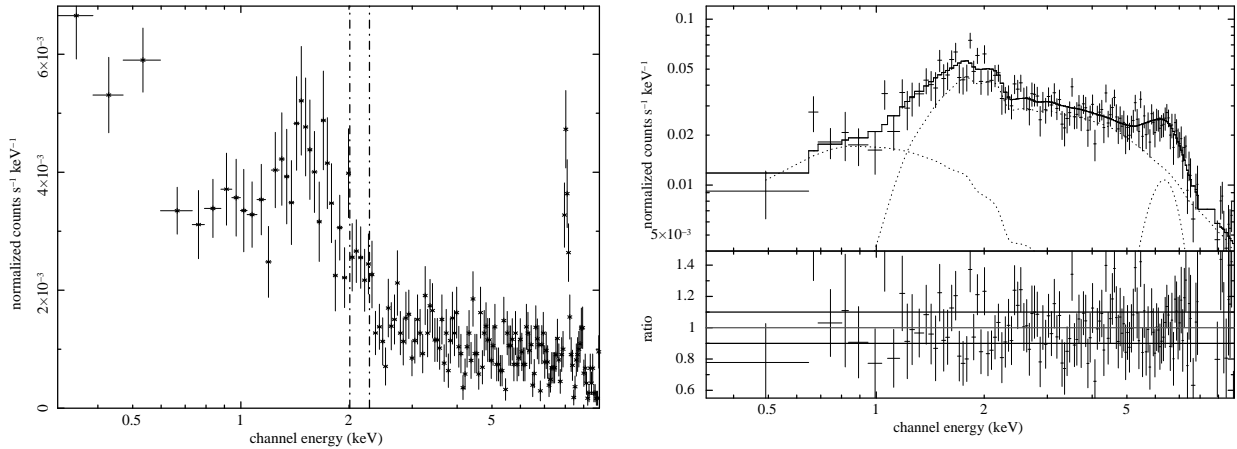

  \centering
  \includegraphics[angle=-90,width=\columnwidth]{background.ps}
  \includegraphics[angle=-90,width=\columnwidth]{halo2.ps}
  \caption{\emph{Left panel}: Background spectrum. Vertical lines show
  the 2.01--2.28 keV energy interval around the absorption feature.
    \emph{Right panel}: Dust scattered halo spectrum.
    }
  \label{skydust}
\end{figure*}

First, we have extracted spectra from two regions containing only the background
and the background plus halo, using the extraction region
described in the previous Section. These
regions are depicted in Fig.~\ref{PNim}, showing the corresponding spectra in
Fig.~\ref{skydust}.
The spectrum of the dust scattered halo area has been corrected for the
background.
We have modelled this spectrum with two absorbed power-law
components assuming a scattered and a soft components as the X-ray
continuum from the halo. We have also included a Gaussian emission
line to describe the fluorescent iron line at 6.4 keV. The residuals
between the spectrum and the model around 2 keV are consistent with
no absorption feature because the ratio differences are smaller
than 10\%.
Therefore, no absorption feature has been detected in the
2.1 keV region either in the halo or in the background itself.
The absorption feature appears only when the source is included in the
extraction region suggesting a local origin.
Therefore, it
must be form either in the neutron star atmosphere or
in the stellar wind of the donor.


On the mechanism of formation of the absorption lines in the isolated neutron star 1E 1207.4$-$5209, the following suggestions have been proposed:

\begin{itemize}
     \item they come from energy level transitions of once ionized helium ions in the strong magnetic
field on the surface of the neutron star (\cite{sanwal02}; \cite{yuan06});
     \item they are electron cyclotron absorption lines in an intense magnetic field
(\cite{sanwal02}; \cite{yuan06});
     \item they are formed by the proton cyclotron absorption in some strong magnetic field
(\cite{sanwal02}; \cite{yuan06});
     \item they are due to atomic transitions in some magnetized iron atmospheres
(\cite{rajagopal}; \cite{mereghetti02}); or
     \item they are due to transitions of hydrogen like O/Ne ions in the stellar atmosphere
with a strong magnetic field (\cite{moha06}).
\end{itemize}

More recently, \cite{xu12} have suggested that the absorption lines in 1E 1207.4$-$5209 could be
explained in the framework of the hydro-cyclotron oscillation model.

On the mechanism of formation of the absorption lines in X-ray binaries, the following suggestions
have been proposed:

\begin{itemize}
     \item they come from absorption K-edges (i.e, \cite{goldstein04});
     \item they are electron cyclotron absorption lines in an intense magnetic field
(see references in Sect.~\ref{intro});
     \item they are due to atomic transitions of hydrogen and helium like Fe ions or
other metals (i. e. \cite{sidoli03}; \cite{boirin04}; \cite{sidoli05}; see also
references in Sect.~\ref{intro}).
\end{itemize}

\subsection{On the formation in the atmosphere of the neutron star}
\label{ns}

In 4U 1538--52, assuming the observed feature is intrinsic to the neutron
star, there are two potential ways to produce it in the
neutron star atmosphere: cyclotron lines and atomic transition lines.

\subsubsection{Cyclotron lines}

First we consider that the observed feature is a cyclotron line produced in a strongly
ionised neutron star atmosphere. If we assume that this is electron cyclotron line, the
feature could be interpreted as the fundamental of the electron-cyclotron energy
$E_{ce} = 11.6 \times B12/(1+z)$ keV in a magnetic field $B12 = B/(10^{12} \; G) \sim 0.24$.
However, 4U 1538--52 presents the fundamental electron-cyclotron line at around 21 keV
(\cite{clark90}; \cite{robba}; \cite{jjrr09}) and the first harmonic at around 47 keV
(\cite{jjrr09}). Therefore, the hypothesis that this feature is an electron cyclotron
line does not look plausible.

Alternatively, one can assume that the spectral feature is associated with ion-cyclotron
energies, $E_{ci} = 0.63 (Z/A)B14$ keV, where Z and A are the atomic charge and atomic mass
of the ion, respectively. The surface magnetic field needed for this interpretation is
greater than the magnetic field inferred for the electron-cyclotron lines detected in
this system. Then, this former scenario seems to be unlikely.

\subsubsection{Atomic lines}

The other possibility is that the observed feature is an atomic line formed in the neutron
star atmosphere. Based on works about the structure and spectra of atoms in strong magnetic
fields, mostly in fields below 10$^{13}$ G (e. g. \cite{ruder94}; \cite{moha02}), we can exclude
this feature as emerging from a hydrogen atmosphere because at any magnetic field and any
reasonable redshift,
there is no hydrogen spectral line whose energy
would match the observed one. In fact, the 2.1 keV absorption
feature cannot be produced by hydrogen atoms as the binding
energy of a hydrogen atom never exceeds $\approx$1 keV at
any magnetic field (\cite{sanwal02}; \cite{moha06}).
As a consequence, spectral features greater than 1 keV
suggest non-hydrogenic element atmosphere on the neutron
star surface.
Therefore, one has to invoke heavier elements.
Another possibility is an iron atmosphere at B around 10$^{12}$ G
(\cite{mereghetti02}; \cite{moha06}), but the iron atmosphere should show many more
than the only one observed feature in the X-ray band (\cite{rajagopal}) and an
unreasonable value of gravitational redshift (\cite{moha06}).
The properties of the absorption line are consistent with
an O/Ne atmosphere (He-like Oxigen, Li-like Oxygen) at B around 10$^{12}$ G
(\cite{moha06}). The O/Ne
atmosphere should show other absorption features at lower energies, but the
soft excess of the system prevents their detection.
We extracted the \emph{RGS} spectrum to look for other
absorption lines. However, the level of source counts at below 2 keV was
too low (\cite{RR11}) and we could not
detect other absorption lines to confirm this origin.

\begin{figure}[h!t]
  \centering
  \includegraphics[angle=-90,width=\columnwidth]{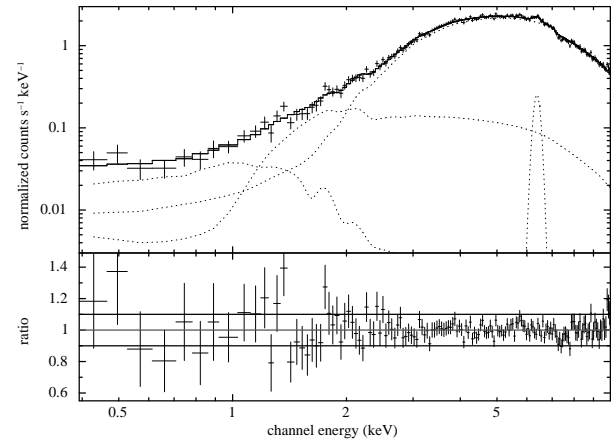}
\vspace{0.1cm}
  \caption{Out-of-eclipse spectrum. The lower panel
  shows the residuals between the spectrum and the best fit
  model. Horizontal lines mark the 10\% ratio differences.
  Soft power-law component
  is changed by a neutron star magnetic atmosphere model.
    }
  \label{nsmax}
\end{figure}

During the out-of-eclipse observation we can see
the neutron star atmosphere directly.
Mid-Z element atmosphere for strongly magnetized neutron
star have been studied by \cite{moho07} (2007). They
constructed spectra with magnetic field of 10$^{12}$ G
and three different effective temperatures for
carbon, oxygen and neon atmospheres. Their models showed numerous
absorption lines, especially in low-temperature models,
presenting heavier element atmospheres more absorption
features (see Figures~11-16 in \cite{moho07} 2007).
Therefore we have interpreted the soft energy
spectrum as generated in a partially ionized, strongly
magnetized mid-Z element plasma (NSMAX model in \emph{XSPEC},
\cite{moho07} 2007, \cite{HPC08} 2008). We have changed the soft
power-law component by a neutron star magnetic atmosphere model.
The absorption feature at 2.1 keV could be produced by
an oxygen/neon atmosphere with B=10$^{12}$ G and effective
temperatures of (3-5)$\times 10^{12}$ K (see Figures~13
and 15 in \cite{moho07} 2007). The magnetic field
derived from the fundamental
cyclotron line of this system $\approx 2.4 \times 10^{12}$ G
is consistent with spectra from magnetized O/Ne
atmosphere models.
Figure~\ref{nsmax} shows the spectra an residuals for this
fit.

\subsection{On the formation in the stellar wind}
\label{wind}

4U 1538$-$52 consists in an accreting neutron star deeply embedded in
the wind of the B0 I star QV Nor. When the X-ray source is observed
through the stellar wind captured by the compact object, absorption K-edges,
such as O, Si, S, and K-and L-edges of Fe are seen (\cite{HWK89}). Many
of them are not apparent in eclipse and the K-edge of Fe was the only
one detected (Paper I). The soft excess at lower energies
hides the possible absorption edges.

We also tested the possibility that the soft
component was due to the presence
of an ionized absorber. Adopting the \textsc{Absori} model in \textsc{xspec}
(\cite{done92}; \cite{arnaud}), we obtained an unsatisfactory fit to the data
and there were still strong negative residuals around 2.1 keV.
\begin{figure*}[h!t]
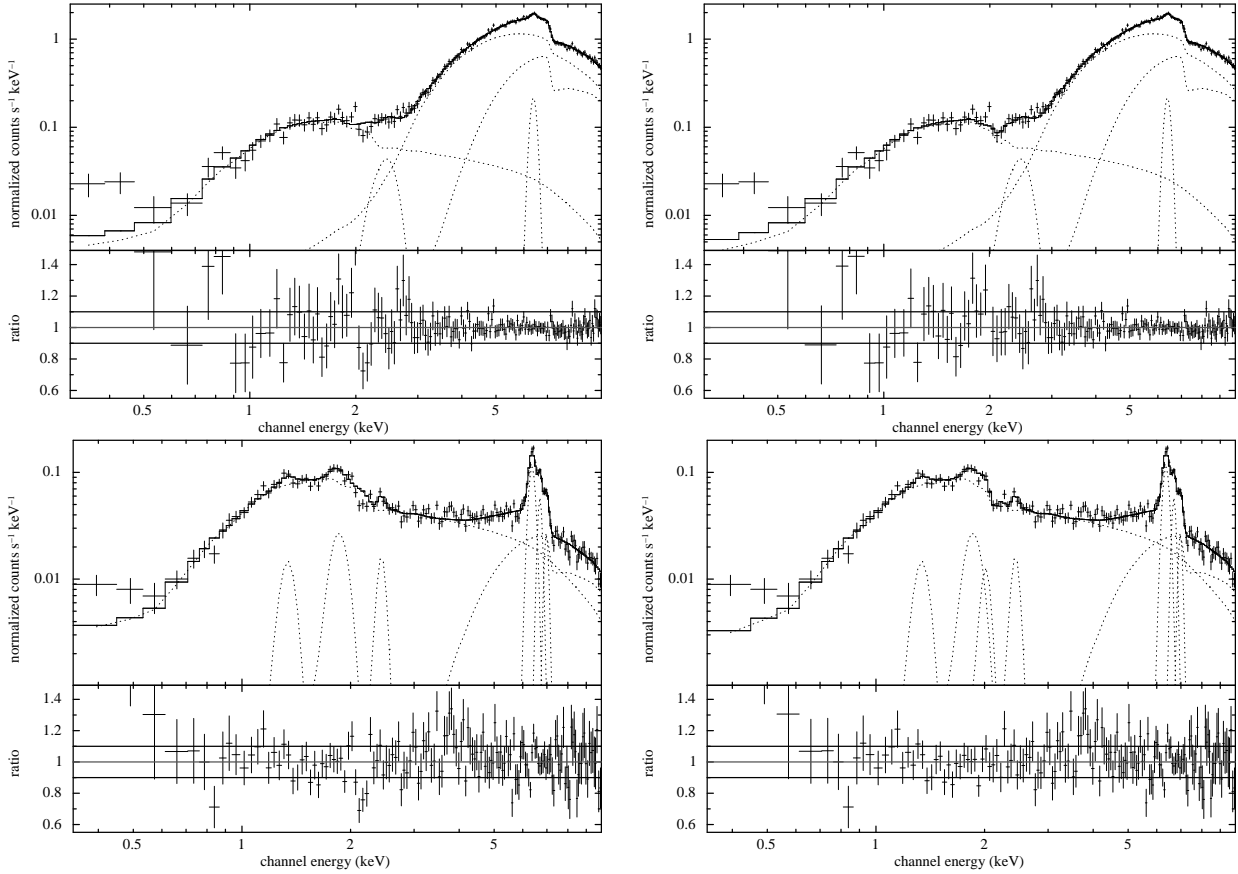

  \centering
  \includegraphics[angle=-90,width=\columnwidth]{pre-eclipse.ps}
  \includegraphics[angle=-90,width=\columnwidth]{pre-eclipse-gabs.ps}
  \includegraphics[angle=-90,width=\columnwidth]{eclipse.ps}
  \includegraphics[angle=-90,width=\columnwidth]{eclipse-gabs.ps}
  \caption{\emph{Top panel}: Eclipse ingress spectrum, best-fit
  model without (left) and with absorption feature (right), and
  the ratio of data to model.
  \emph{Bottom panel}: Eclipse spectrum, best-fit model
    without (left) and with absorption feature (right),
    and the ratio of data to model.
    The absorption line is clearly seen.
    }
  \label{source}
\end{figure*}

Figure~\ref{source} shows the eclipse ingress spectrum and
best-fit model (three absorbed power-laws modified by two
Gaussian emission lines) in the 0.3--10.0 keV energy range
obtained with \emph{PN} camera and residuals between the spectrum
and the model (top left panel), the eclipse spectrum
and best-fit model (two absorbed power-laws modified
by six Gaussian emission lines) and residuals between
the spectrum and the model (bottom left panel).
Assuming the absorption feature is formed in the
stellar wind we have fitted the best-fit model derived
in this work to the pre- and
eclipse spectra without and with a Gaussian absorption line
at 2.1 keV (see Figure~\ref{source} top right and bottom
right panels, respectively). 
Adding
a Gaussian absorption line both in the hard and in the scattered
power-law component we obtained no improvement in the fit.
But adding it in the soft power-law component we obtained a significant improvement in the fit of the eclipse spectrum (f-test probability
1.2$\times 10^{-6}$) and a slightly improvement in the fit
of the eclipse ingress spectrum (f-test probability 0.020). The best-fit 
parameters of the line at 2.1 keV are given in
Table~\ref{abslieclparam}. Uncertainties refer to a single parameter
at 90\% ($\Delta\chi^2=$2.71) confidence limit.
\begin{table}
\begin{minipage}[t]{\columnwidth}
  \caption{Fitted parameters for the absorption line detected
  in the eclipse ingress and the eclipse spectrum (Fig.~\ref{source}).
    }
\label{abslieclparam}
\centering                          
\renewcommand{\footnoterule}{}      
\begin{tabular}{r r c}        
\hline                 
Component & Parameter & Value \\    
\hline                        
\multicolumn{3}{c}{Eclipse ingress spectrum} \\
\hline
\emph{Absorption} & E (keV) & 2.13$^{+0.05}_{-0.04}$ \\      
\emph{line} & $\sigma$ (eV) & 7$\pm$6 \\
 & tau & 1.0 (unconstrained) \\
 & EW(eV) & 30$\pm$16 \\
 & & \\
 & $\chi^2_\nu$(dof) & 1.0(157) \\ 
\hline                                   
\multicolumn{3}{c}{Eclipse spectrum} \\
\hline
\emph{Absorption} & E (keV) & 2.12$\pm$0.03 \\      
\emph{line} & $\sigma$ (eV) & 9$^{+4}_{-5}$ \\
 & tau & 1.0 (unconstrained) \\
 & EW(eV) & 39$^{+11}_{-21}$ \\
 & & \\
 & $\chi^2_\nu$(dof) & 1.3(154) \\ 
\hline                                   
\end{tabular}
\end{minipage}
\end{table}

In order to identify the possible ions from which this feature could
originate, we used atomic database available such as the van Hoof's
Atomic Line List\footnote{http://www.pa.uky.edu/$\sim$peter/atomic/index.html},
the X-ray transition energies from the National Institute of Standards and
Technology (NIST)\footnote{http://www.nist.gov/pml/data/xraytrans/index.cfm},
the line finding list from XSTAR
package\footnote{http://heasarc.gsfc.nasa.gov/docs/software/xstar/xstar.html}
and the CHIANTI atomic database for spectroscopic diagnostics of astrophysical
plasmas\footnote{http://www.www.chianti.rl.ac.uk/line-list.html} (\cite{dere97}; \cite{dere09}).
Therefore, we looked for X-ray transitions in the energy range 2.03--2.21 keV
or in the wavelength range 5.61--6.11 \AA.

\begin{table}
\begin{minipage}[ht]{\columnwidth}
  \caption{X-ray transitions identified from the atomic databases used in this work.
    }
\label{lines}
\centering                          
\renewcommand{\footnoterule}{}      
\begin{tabular}{l l c l}        
\hline         
Element/Ion & Transition & E & $\lambda$ \\    
& & (keV) & (\AA) \\
\hline                        
\multicolumn{4}{c}{van Hoof's atomic line list} \\
Fe \textsc{xxvi} & \tiny{2--9} & 2.202 & 5.632 \\
Si \textsc{xiii} & \tiny{1S--1P$^\circ$} & 2.183 & 5.681 \\
Fe \textsc{xxvi} & \tiny{2--8} & 2.171 & 5.711 \\
Fe \textsc{xxvi} & \tiny{2--7} & 2.127 & 5.829 \\ \hline
 & & \\
\multicolumn{4}{c}{NIST's database} \\
Si \textsc{xiii} & \tiny{1S--1P$^\circ$} & 2.183 & 5.681 \\ \hline
 & & \\
\multicolumn{4}{c}{CHIANTI's database} \\
Si \textsc{xii} d & \tiny{1s$^2$ 2p $^2$P$_{1/2}$--1s 2p ($^3$P) 3p $^2$D$_{3/2}$} & 2.1322 & 5.8156 \\
Si \textsc{xii} d & \tiny{1s$^2$ 2p $^2$P$_{3/2}$--1s 2p ($^3$P) 3p $^2$D$_{5/2}$} & 2.1320 & 5.8162 \\ \hline
 & & \\
\multicolumn{4}{c}{Inner-shell absorption line (\cite{hullac})} \\
Si He \textsc{i} & \tiny{1S--3P} & 2.182 & 5.682 \\
\hline                                   
\end{tabular}
\end{minipage}
\end{table}
Nevertheless, we note that the cosmic abundance for
silicon is $3.55\times10^{-5}$ while for iron, aluminium, nickel, phosphorus, and zinc is
$4.68\times10^{-5}$, $2.95\times10^{-6}$, $1.78\times10^{-6}$, $2.82\times10^{-7}$, and
$3.98\times10^{-8}$, respectively. Taking this into account, we found
the allowed transitions which are listed in Table~\ref{lines} and suggest that the absorption
feature is related to silicon (Si \textsc{xii} or Si He \textsc{i}) or iron (Fe \textsc{xxvi})
ions.

\emph{BeppoSAX} spectrum of the LMXB EXO 0748$-$676 showed an absorption feature at
2.13 keV which was tentatively identified with absorption from highly ionized silicon
(Si \textsc{xiii}, \cite{sidoli05}). However, they did not performed any instrumental analysis in order
to discard an inadequate modelling of the instrumental response.

\section{Conclusion}

We have presented the spectral analysis of the HMXB 4U 1538$-$52 using
the \emph{XMM-Newton} observation focusing our attention specifically
in the absorption feature at 2.1 keV. In this work, we show that the \emph{XMM} spectrum of 4U 1538$-$52 shows a deep significant
absorption feature at E = 2.12$\pm$0.03 keV
in the eclipse spectrum. Throughout a detailed analysis of the spectral
region covering the absorption line, we have been able to discard an instrumental origin.
Indeed, it is not present in any of the long term monitoring sources used for the \emph{XMM}
calibration. Likewise, it can not be accounted for by gain and/or offset corrections and the
latest calibration files. We conclude that the line is of astrophysical origin.

In order to constrain its nature we have analysed the spectra from the dust scattered halo
seen around the source and also the spectrum of the background itself. The absorption line is
not present in either of the two. This strongly suggests that the line is formed locally, in
the binary system either in the stellar wind or in the atmosphere of the
neutron star. We found two possible physical scenarios where the absorption
line could be formed: an O/Ne atmosphere on the neutron star surface
or atomic transitions of hydrogen
and helium like Fe or Si ions in the stellar wind of the donor.
We have not discarded its formation in
the atmosphere of the neutron star, assuming it could be seen during
the out-of-eclipse observation. Moreover, the 2.1 keV
feature present in our eclipse spectrum has $\approx$8 times higher
equivalent width and deviations from the continuum model
at the level of $\approx$25\%. We also notice that
for on axis sources \emph{EPIC} calibration accuracy is better than 5\%.
These results give us confidence that
the feature is intrinsic to the system.
Consequently, to confirm the nature of the absorption line
one should investigate the presence of other features in the
low energy band with a high-resolution and enough exposure time
to achieve a good signal-to-noise at low energies.

Either scenario is of great astrophysical interest. We
tried to look for more absorption lines analysing the \emph{RGS}
data of this source, but the level of counts was compatible
with the background because this system is highly absorbed
at low energies. Further studies
at high resolution will be needed to use this absorption line
as a potential diagnostic tool to study the properties of the neutron star
atmosphere or the stellar wind of the donor.

\acknowledgements
  We are grateful to the anonymous referee for
  useful and detailed comments.
  This work was supported by the Spanish Ministry of Education and Science
  \emph{De INTEGRAL a IXO: binarias de rayos X y estrellas activas} project
  number AYA2010-15431.
  This research has made use of data
  obtained through the XMM-Newton Science Archive (XSA),
  provided by European Space Agency (ESA) and Peter van Hoof's
  Atomic Line List (http://www.pa.uky.edu/$\sim$peter/atomic/index.html).
  CHIANTI is a collaborative project involving researchers at NRL (USA) RAL (UK),
  and the Universities of: Cambridge (UK), George Mason (USA), and Florence (Italy).
  We would like to thank
  the \emph{XMM helpdesk}, particularly Matteo Guainazzi, for
  invaluable assistance in determining the systematic
  uncertainties in the \emph{PN} data.
  JJRR acknowledges the support by the
  Spanish Ministerio de Educaci\'on y Ciencia under grant
  PR2009-0455 and by the Vicerectorat d'Investigaci\'o,
  Desenvolupament i Innovaci\'o de la Universitat d'Alacant
  under grant GRE12-35.

\newpage

\end{document}